%% file: ms.tex
\documentclass[apj]{emulateapj}
\usepackage{apjfonts} 

\usepackage{epsfig}
\usepackage{longtable}
\usepackage{graphicx}

\newcommand{\n}{\nodata}

\def\bi{\begin{itemize}}
\def\ei{\end{itemize}}
\def\be{\begin{equation}}
\def\ee{\end{equation}}

\def\gtrsim{\mathrel{\hbox{\rlap{\hbox{\lower4pt\hbox{$\sim$}}}\hbox{$>$}}}}
\def\lesssim{\mathrel{\hbox{\rlap{\hbox{\lower4pt\hbox{$\sim$}}}\hbox{$<$}}}}
\def\gtrsim{\mathrel{\hbox{\rlap{\hbox{\lower4pt\hbox{$\sim$}}}\hbox{$>$}}}}
\def\lesssim{\mathrel{\hbox{\rlap{\hbox{\lower4pt\hbox{$\sim$}}}\hbox{$<$}}}}

\shortauthors{Lister et al.}
\shorttitle{MOJAVE. V. Multi-epoch VLBA Images}

\received{2008 October 24}
\accepted{2008 December 19}
\journalinfo{Astronomical~journal}
\submitted{}

\begin{document}
\title{MOJAVE: Monitoring of Jets in AGN with VLBA Experiments. V. Multi-epoch VLBA Images}

\author{M. L. Lister}
\affil{Department of Physics, Purdue University, 525 Northwestern
Avenue, West Lafayette, IN 47907; mlister@purdue.edu}

\author{H. D. Aller, M. F. Aller}
\affil{Department of Astronomy, University of Michigan, 817 Denison Building, Ann Arbor, MI 48109-1042, USA; haller@umich.edu, mfa@umich.edu}

\author{M. H. Cohen}
\affil{Department of Astronomy, California Institute of Technology, Mail Stop 105-24, Pasadena, CA 91125, USA; mhc@astro.caltech.edu}

\author{D. C. Homan}
\affil{Department of Physics and Astronomy, Denison University,
Granville, OH 43023; homand@denison.edu}

\author{M. Kadler}

\affil{1)Dr. Remeis-Sternwarte Bamberg, Universit\"at Erlangen-N\"urnberg, Sternwartstrasse 7, 96049 Bamberg, Germany; 2) Erlangen Centre for Astroparticle Physics, Erwin-Rommel Str. 1, 91058 Erlangen, Germany; 3) CRESST/NASA Goddard Space Flight Center, Greenbelt, MD 20771, USA; 4) Universities Space Research Association, 10211 Wincopin Circle, Suite 500 Columbia, MD 21044, USA; matthias.kadler@sternwarte.uni-erlangen.de}

\author{K. I. Kellermann}
\affil{National Radio Astronomy Observatory, 520 Edgemont Road,
Charlottesville, VA 22903-2475; kkellerm@nrao.edu}

\author{Y. Y. Kovalev}
\affil{Max-Planck-Institut f\"ur Radioastronomie, Auf dem H\"ugel 69,
D-53121 Bonn, Germany and Astro Space Center of Lebedev Physical Institute,
Profsoyuznaya 84/32, 117997 Moscow, Russia; ykovalev@mpifr-bonn.mpg.de}

\author{E. Ros, T. Savolainen, J. A. Zensus}
\affil{Max-Planck-Institut f\"ur Radioastronomie, Auf dem H\"ugel 69,
D-53121 Bonn, Germany;\\ ros@mpifr-bonn.mpg.de, tsavolainen@mpifr-bonn.mpg.de, azensus@mpifr-bonn.mpg.de,}

\author{R. C. Vermeulen}
\affil{ASTRON, Postbus 2, NL-7990 AA Dwingeloo, Netherlands; rvermeulen@astron.nl}

\begin{abstract}
We present images from a long term program (MOJAVE: Monitoring of Jets
in AGN with VLBA Experiments) to survey the structure and evolution
of parsec-scale jet phenomena associated with bright radio-loud active
galaxies in the northern sky. The observations consist of 2424 15 GHz
VLBA images of a complete flux-density limited sample of 135 AGN above
declination $-$20 degrees, spanning the period 1994 August to 2007
September. These data were acquired as part of the MOJAVE and 2~cm
Survey programs, and from the VLBA archive.  The sample selection
criteria are based on multi-epoch parsec-scale (VLBA) flux density,
and heavily favor highly variable and compact blazars. The sample
includes nearly all the most prominent blazars in the northern sky,
and is well-suited for statistical analysis and comparison with
studies at other wavelengths. Our multi-epoch and stacked-epoch images
show 94\% of the sample to have apparent one-sided jet morphologies,
most likely due to the effects of relativistic beaming. Of the
remaining sources, five have two-sided parsec-scale jets, and three
are effectively unresolved by the VLBA at 15 GHz, with essentially all
of the flux density contained within a few tenths of a milliarcsecond.

\end{abstract}

\section{Introduction}

High angular resolution studies of the kinematics of AGN jets over
many years have led to a better understanding of the process of
acceleration and collimation of the relativistic jets associated with
the ejection of relativistic plasma from a central supermassive black
hole.  Early observations were made with ad hoc arrays comprised of
relatively few antennas and focused on a few relatively strong and
fast superluminal sources (e.g., \citealt*{C75, S75, SCM75}).  In
these studies it was difficult to trace details of the jet outflow due
to limited temporal and interferometric coverage of the arrays,
which consisted of existing antennas with differing instrumental
characteristics. Moreover, by concentrating on only the fastest or
most interesting sources, (e.g., \citealt*{C77,ZCU95}), the kinematic
results did not reflect the full range of speeds seen in the overall
population. With the start of VLBA operations in the mid 1990s, we
were able to obtain a reasonably uniform set of multi-epoch 15 GHz (2
cm) observations on 132 AGN \citep{K98,2002AJ....124..662Z} which covered the period
1994 to 2002. From these data we were able to discuss the statistical
properties of AGN jets \citep {KL04,KKL05,H06}, including the
intrinsic properties of the parent population
\citep{CLH07} and the detailed nature of a few sources of particular
interest \citep{H03,VRK03,LKV03,K08,KLH07}.

Although these studies included most known ``flat spectrum'' ($\alpha
> -0.5$, where $S_\nu \propto \nu^\alpha$) radio sources with total
flux density above well defined limits, they included or excluded some
sources because of the contribution of extended emission, errors in
the measured spectral index and/or spectral curvature.  As described
in paper I of this series \citep{LH05}, our new MOJAVE (Monitoring of
Jets in Active Galactic Nuclei with VLBA Experiments) sample includes
all sources with measured total VLBA flux density at 15 GHz greater
than a defined flux density at any epoch during the period 1994.0 to
2003.0.  For the 96 sources in common with the studies of \cite{K98} and 
\cite{2002AJ....124..662Z}, the time scale for determining the speed and possible
accelerations or decelerations is now extended from 7 to 13
years. \cite{LH05} give more details of the sample definition and also
discuss the single-epoch linear polarization properties of the MOJAVE
sample, while \cite{HL06} discuss the circular polarization properties
(Paper II).

The MOJAVE/2 cm Survey program is among the largest multi-epoch AGN
VLBI surveys carried out to date, and is complemented by several programs
at other wavelengths. These include the 5 GHz Caltech-Jodrell
Flat-Spectrum Survey \citep{Britzen07,2008AA...484..119B}, the 2 \& 8 GHz Radio Reference
Frame program \citep{PMF07}, and the 43 GHz Boston University AGN
monitoring program \citep{J07}. Each of these programs offers various
trade-offs in terms of sample size, angular resolution, and image
sensitivity. Shorter wavelength studies, such as the Boston University
program, provide better angular resolution than at 15 GHz, but the
sensitivity is poorer, jet features fade faster, and the impact of
tropospheric conditions  is more severe.  Thus, short wavelength
observations need to be more frequent, and consequently must focus on much
smaller samples of extremely bright sources.  Longer wavelength studies have
better surface brightness sensitivity and so can trace the jet motions
out to greater distances, but at the expense of degraded angular resolution.

This paper is the fifth in a series describing the results of the
MOJAVE program, with the most recent ones discussing the kiloparsec-scale jet properties
\citep{CLK07}(Paper III) and parent luminosity function
\citep{CL08}(Paper IV) of the sample. Here we 
present a complete set of 15 GHz VLBA images of the flux-density
limited MOJAVE sample of 135 compact extragalactic radio jets obtained
during the period 1994 August to 2007 September. The overall layout of
the paper is as follows: in \S~\ref{SD} we discuss the sample
definition and selection criteria, in \S~\ref{obs} we discuss our
observational program and data reduction method, and in \S~\ref{morph}
we summarize the overall parsec-scale morphological properties of the
AGN jets and describe our current monitoring program.  In subsequent
papers we will discuss their kinematic properties, overall
demographics, and polarization properties.

\section{Sample Definition \label{SD}} 

In Paper I we described our selection criteria for the MOJAVE sample,
which is based on compact radio flux density in order to provide the
best comparison with Monte Carlo samples of relativistically beamed
jets (e.g., \citealt*{LM97}).  By using the milliarcsecond scale
(VLBA) 15 GHz flux density rather than total flux density as the
selection criterion, we were effectively able to exclude any
contribution from large-scale emission, leaving a sample consisting
almost entirely of radio loud AGN with relativistic jets pointed close
to the line of sight (i.e., blazars). The only exceptions were a
handful of nearby radio galaxies and peaked-spectrum sources whose jet
axes probably lie much closer to the plane of the sky.

The MOJAVE selection criteria resemble those of the
original VLBA 2 cm Survey of \cite{K98}, and are defined as follows:
\begin{itemize}
\item{} J2000 declination  $\ge -20 \arcdeg$
\item{} Galactic latitude $|b| \ge 2.5 \arcdeg$
\item{} Total 15 GHz VLBA flux density of at least $1.5$ Jy ($\ge 2$ Jy for sources below the
   celestial equator) {\it at any epoch} during the period 1994.0--2004.0.
\end{itemize}
Because of the highly variable nature of strong, compact AGN, we did
not limit the flux density criterion to a single fixed epoch. Doing so
would have excluded many highly variable sources from the sample, and
would subsequently reduce the robustness of statistical tests on
source properties. Including the variable AGN also provides a more
complete sample for comparisons with AGN surveys at other wavelengths,
such as the 3rd EGRET \citep{HBB99} and upcoming Fermi
\citep{Thomp04} gamma-ray catalogs. The higher flux density limit for the
southern sources was chosen since the VLBA has reduced hour angle
coverage for this region of the sky and consequently poorer imaging
capability. 

We note that for several sources in our sample, we did not have any 15
GHz VLBA measurements that formally met our flux density criterion. However, we
were able to infer their VLBA flux density at other epochs by using an
extensive database of flux density measurements from the UMRAO and
RATAN telescopes spanning 1994.0 to 2004.0 (e.g.,
\citealt*{Kovalev_etal99,AAH92}). An important step in this method was
to determine the amount of extended (non-VLBA) flux density in each
source using near-simultaneous UMRAO/VLBA measurements made at multiple
epochs.

In Paper I we originally identified a total of 133 AGN that satisfied
these criteria. In processing additional archival VLBA epochs for the
current paper, we have subsequently identified two more sources
(0838+133 and 1807+698) which met our selection criteria, bringing the
total number up to 135 AGN. One additional source, 1150+497, appears
to have possibly met the MOJAVE criteria based on interpolation of
single-dish RATAN radio observations in 2003 August, however, we did
not obtain any 15 GHz VLBA epochs on this source prior to 2008. This
source is now currently in our extended monitoring program (see
\S~\ref{mojave2}). We believe the overall completeness of the MOJAVE survey to be 
high, with few sources missed, because of our large pre-selection
candidate list (see Paper I) and the extensive flux density database that
was available from the VLBA, RATAN, and UMRAO facilities.

We list the general properties of the MOJAVE sources in Table~1.  The
redshifts are gathered from the literature, and are currently only
93\% complete, mainly because of the featureless spectra of several BL
Lacs in the sample.  We list the nature of the optical counterpart
from \cite{VCV06} for each AGN in column 6, except as noted in Table
1. At the present time, only four sources (0446+113, 0648$-$165,
1213$-$172, and 2021+317) lack published optical counterparts.

The MOJAVE sample is dominated by quasars, with the weak-lined BL
Lacertae objects and radio-galaxies making up 16\% and 6\% of the
sample, respectively. \cite{VCV06} have recently reclassified as
quasars some BL Lac objects that have occasionally displayed emission
lines slightly wider than the nominal 5 \AA $\,$ equivalent width
limit \citep{SFK91}. Since these are still much narrower than those
typically found in quasars, we have chosen to list the previous BL Lac
classifications for these objects, and have indicated these with flags
in Table 1. The classifications of the radio spectral shape given in
column 7 are described by \cite{KL04}. The MOJAVE sample is heavily
dominated by blazars with flat overall radio spectra, which we define
as a spectral index $\alpha$ flatter than $-0.5$ at any frequency
between 0.6 and 22 GHz.  The five steep-spectrum sources in the sample
have strong extended emission on arcsecond scales that dominates the
integrated spectrum, but their parsec-scale emission still met our
selection criteria. Two of the latter AGN (1458+718 and 1823+568)
are frequently referred to as compact steep spectrum sources in the
literature. All of the sources originally described in Paper I as
peaked-spectrum, with the exception of 0742+103, have subsequently
turned out to have variable radio spectra and are therefore no longer
considered as stable peaked-spectrum sources (e.g.,
\citealt*{Torn05}).

\section{Observations and Data Reduction }
\label{obs}

The 15 GHz VLBA data presented in this paper consist of observations
carried out between 1994 and 2007 as part of the VLBA 2 cm Survey and
MOJAVE programs, and data from the online NRAO data
archive\footnote{http://archive.cv.nrao.edu/} spanning the period 1998
-- 2006. The multi-epoch observations for each source, along with the
corresponding image parameters, are described in Table 2.  The 2 cm
Survey observations analyzed by \cite{KL04} and E. Ros et al. (in
preparation) consist of approximately one hour integrations on each
source, broken up into approximately five minute-long scans separated
in hour angle to improve {\it(u,v)} plane coverage. A similar
observing method and integration times were used in the MOJAVE
observations from 2002 May to 2007 September (VLBA codes BL111, BL123,
BL137, and BL149; as flagged in Table 2), and are described in Paper
I. During 2006 (VLBA codes BL137), the 15 GHz integration times were
shortened by a factor of $\sim 3$ to accommodate interleaved scans at
three other observing frequencies (8.1, 8.4, 12.1 GHz). The MOJAVE and
2 cm Survey observations were recorded at a data rate of 128 Mbps,
which was increased to 256 Mbps in the epoch 2007 July 3 and epochs
thereafter. Beginning with the 2007 January 6 epoch, the number of
sources observed in each 24 hour MOJAVE session was increased from 18
to 25 sources to accommodate our larger monitoring sample (see
\S~\ref{mojave2}).

In order to obtain the best possible temporal coverage to study jet
kinematics, we re-analyzed all past VLBA 15 GHz observations that were
available in the VLBA archive from 1998--2006 with good
interferometric data on the sources in our sample.  Since we require
good {\it(u,v)} coverage to track jet motions accurately, we only
considered observations in which there were at least three scans on
the source, suitably separated in hour angle. We also omitted any
archival observations that had fewer than 8 VLBA antennas. Any
observations which lacked data from either of the outermost antennas
(Mauna Kea and St. Croix) were also omitted if they had fewer than 6
scans on the remaining antennas.  Finally, we excluded a few archival
epochs on some sources after processing them if their overall data
quality was very poor due to weather and/or technical
problems. Further details on these archival epochs can be obtained
from the NRAO data archive using the appropriate VLBA code and
observation date listed in Table~2.

Including the MOJAVE and 2 cm Survey epochs, we analyzed a total of
416 VLBA experiments, yielding 2424 individual source epochs on the
MOJAVE sample.  A portion of these were re-analyzed and/or presented
in earlier 2 cm Survey and MOJAVE papers. We present the entire set of
source epochs here for completeness.

The temporal coverage per source varies between 5 and 89 epochs, with
a median of 15 epochs (column 5 of Table~1). The wide dispersion
reflects the fact that some AGN are observed much more frequently with
the VLBA, either as calibrators, or because they are sources of
particular astrophysical interest. Also, attempts are made in the 2
cm Survey and MOJAVE programs to observe the sources with the most
rapid motions more frequently, given the constraints associated with
VLBA's dynamical scheduling system. In setting the observing
intervals, we have aimed to not let the jet features move more than
about half of their typical angular spacing between epochs. This is
done to facilitate the unambiguous identification of the same features
from one epoch to the next.

\subsection{Data Reduction \label{data_reduct}}

We processed the data according to the standard procedures described
in the AIPS cookbook\footnote{http://www.aips.nrao.edu}. Our overall
reduction method was to initially flag out bad data on the basis of
excessively high system temperature (typically due to inclement
weather), remove sampler bias with the task ACCOR, calibrate the
correlator output using antenna system temperature and gain curves,
and apply an atmospheric opacity correction with APCAL.  The pulse
calibration signals were used in all the experiments to align the
phases across the IFs. In a few cases where no pulse calibration
signal was available, we performed a fringe fit of a single scan on a
bright calibrator and used the result to align the delays for the
other sources.  In the case of the archival epochs, and for MOJAVE
epochs observed after 2007 Aug 9, we deemed the standard global
fringe-fitting step to be unnecessary because of the strong and
compact nature of all the sources, their well-determined sky positions
(e.g., from the ICRF, RDV, and VCS surveys, e.g.,
\citealt*{icrf-ext2,VCS6}), and the well-determined positions of the
VLBA antennas.  After a final bandpass correction and a short (30
second) phase self calibration using a point source model, we exported
the data to Difmap \citep{difmap}. The imaging and self-calibration
process in Difmap was largely automated, with human input required
mainly for data flagging. The procedure was facilitated by the use of
a {\it mySQL} database containing the locations of clean boxes for
each source, which we determined during the processing of previous
epochs. Since the overall extent of the individual sources did not
significantly vary over the time period covered by our survey, the use
of pre-determined clean boxes provided additional a-priori information
that guided the self-calibration process and facilitated rapid
convergence toward the final model. For experiments in which
polarimetric cross-hand data were recorded, we performed additional
calibration steps to correct for antenna leakage terms and the
absolute electric vector position angle on the sky, as described in
paper I.  These multi-epoch polarization data will be presented in a
separate paper.

\subsection{Images \label{images}}

We produced the final images using Difmap with a multi-resolution
clean algorithm developed by \cite{Moe98}. The technique involves
`super-resolving' the brightest and most compact features in the
initial cleaning stages, which leads to a more accurate representation
of lower surface brightness features in the final (natural-weighted)
cleaning stages. The final step consisted of a large iteration deep
clean, which also helps bring out these features. This procedure does,
however, alter the noise statistics in the final cleaned image.  For
this reason we obtained the rms image noise values from a blank sky
region a full arcsecond away from the CLEAN image center after the
final cleaning stage. This provided a better estimate of the true rms
image noise associated with the interferometric visibilities, since
this region was located well outside the deep CLEAN box of the main
image.

We present the un-tapered, naturally-weighted images of all the
individual source epochs in Figure Set 1.  The dimensions of the
restoring beam in each image vary depending on the {\it(u,v)} coverage
of the individual observation. Information on the restoring beam can
be found in Table~2, along with the lowest contour level, VLBA
experiment code, and observation date. The image dynamic range varies
with {\it(u,v)} coverage and integration time, with typical values of
approximately 5000:1.  We obtained the total flux densities by summing
all emission from pixels in the image above the lowest contour level
within the region shown in Figure~1. Gaussian model fits to the
visibility data, as well as a kinematic analysis of the sample, will
be presented in a separate paper.

\begin{figure*}
\centering
\resizebox{1.0\hsize}{!}{
   \includegraphics[trim=0.8cm 1cm 0.8cm 2cm]{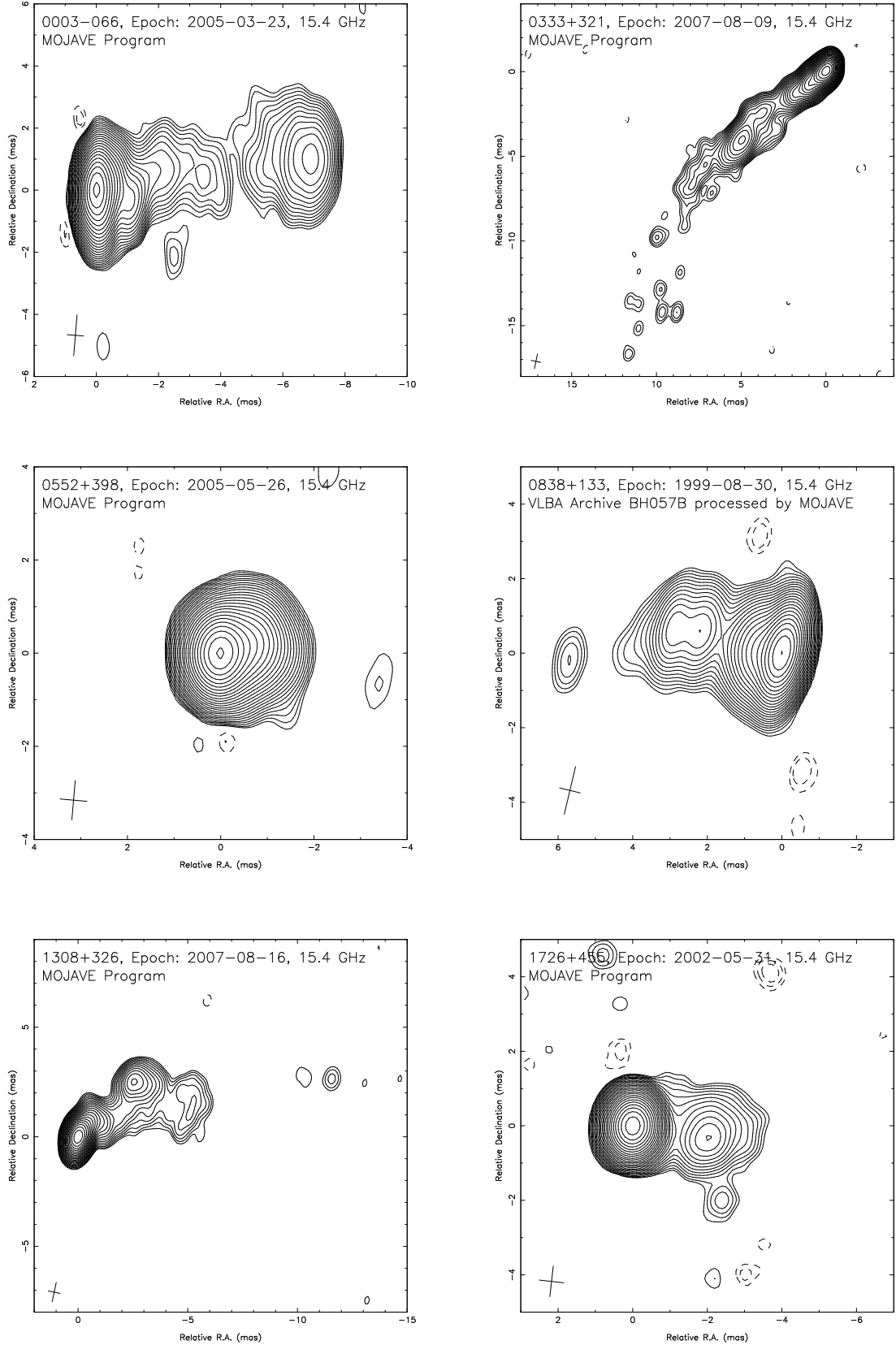}
}
\caption{\label{f:images} Naturally-weighted 15 GHz total intensity VLBA contour images
of individual epoch observations of the MOJAVE AGN sample. The
contours are in successive powers of $\sqrt{2}$ times the base contour
level given in Table 2. Because of self-calibration, in some cases the
origin may be coincident with the brightest feature in the image,
rather than the putative core component listed in Table~3. Clockwise
from the top left: Image of 0003$-$066, 0333+321, 0838+133, 1726+455,
1308+326, 0552+398.  Note: this is a figure stub, the full figure set
is available in the electronic version.
}
\end{figure*}

\begin{figure*}
\centering
\resizebox{1.0\hsize}{!}{
   \includegraphics[trim=0.8cm 1cm 0.8cm 2cm]{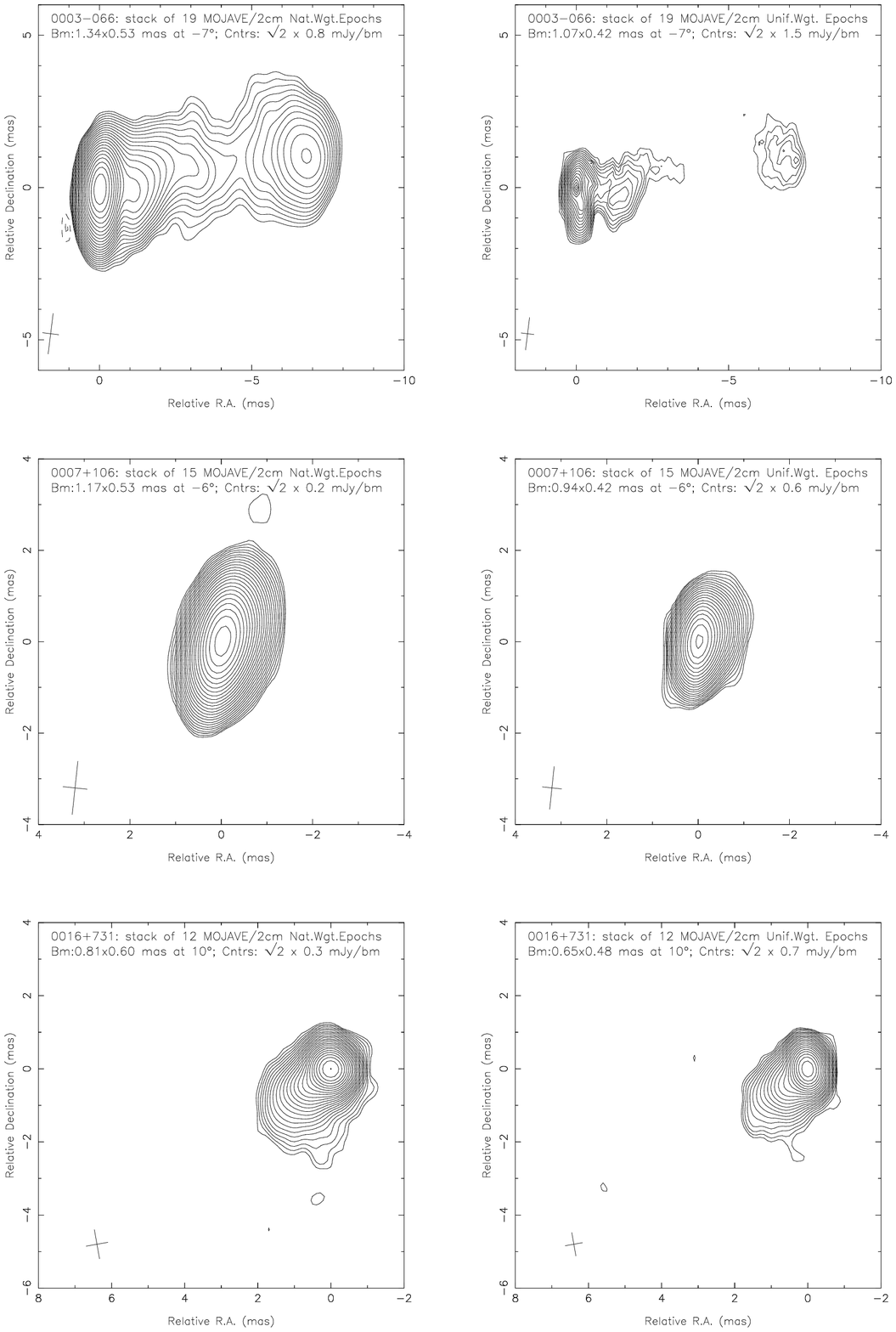}
}
\caption{\label{f:stackedimages} Total intensity 15 GHz
VLBA stacked-epoch contour images of the MOJAVE AGN sample. The left
hand panels represent the unweighted averages of the naturally
weighted images presented in Figure Set 1, after shifting each map to
make the core component coincident with the origin. The right hand
panels contain stacked uniformly-weighted images with beam dimensions
80\% of the naturally weighted images. From top to bottom are stacked
images of 0003$-$066, 0007+106, and 0016+731. Note: this is a figure stub, the full figure set
is available in the electronic version.
}
\end{figure*}

In order to show the fullest possible extent of the jets, we have also
produced `stacked' naturally- and uniformly-weighted images of each
source (Figure Set 2). We created the former by first restoring the image
at each epoch with an identical beam whose dimensions corresponded to
the median naturally-weighted beam of all epochs on that source. We
then shifted the convolved image at each epoch to align the fitted
positions of the core features, and then combined them all with equal
weight to produce an averaged (stacked) image. We created the
uniformly-weighted images in the same manner, using a restoring beam
that was 80\% of the naturally weighted beam size in both dimensions.

We note that in the case of a few sources located within $\sim 1.5$
degrees of the celestial equator, the individual baseline tracks in
the {\it(u,v)} plane are almost exclusively in the east-west
direction, resulting in relatively poorer interferometric
coverage. This can create artifacts to the direct north and south of
the brightest feature, as seen in some of the stacked and individual
epoch images (e.g., 1055+018, 2134+004). VLBA images of equatorial
sources with jets aligned in the north and south directions on the sky
suffer particularly from this effect.

A very small number of our images may also suffer from the effects of
rapid intraday variability (IDV). Since the imaging procedure assumes
the source flux density is not significantly changing on the (several
hour long) timescale of the observation, intrinsic source variability
can add error to the gain solutions, with a resulting increase in
image noise. We have discovered one such case in 1156+295, epoch 2007
Feb 5 \citep{SK08}. We are currently analyzing our whole dataset for
other instances of IDV (Kuchibhotla et al., in preparation).


The full set of all images are included in the online version of this
article. Linearly interpolated movies which show the jet motions and
structural variability, FITS images, and calibrated visibility data
are also publicly available on the MOJAVE
website\footnote{http://www.physics.purdue.edu/MOJAVE}. The images
presented in this paper cover epochs up to the end of 2007. More
recent images and calibrated visibility data from our current program
are uploaded to the web site as soon as they are reduced, typically 3
to 4 weeks after each observing session. All of the raw visibility
data associated with the MOJAVE program are publicly available
immediately after correlation from the NRAO data archive.

\section{Morphology}
\label{morph}

The appearance of blazar jets depends not only on their intrinsic
structure, but also on the Doppler factor associated with the highly
relativistic motion of individual features in the jet. Their small
viewing angles can highly exaggerate mild curvatures in the jet flow,
and relativistic beaming effects can amplify or diminish the apparent
brightness of different jet regions as the result of even minor
changes in the direction of motion with respect to the line of sight.
It is thought that all jets must be intrinsically double-sided to
explain the transport of energy from the central engine to the double
radio lobes commonly found in FRI and FRII radio galaxies and in some
quasars. In many cases, however, the visible morphology of these jets
is highly affected by differential Doppler boosting. The apparent flux
density ratio of the approaching and receding jets may reach as high
as $10^6$ in the fastest known jets, giving the illusion of a
one-sided jet in VLBI images made with limited sensitivity and dynamic
range.

All but eight MOJAVE sources in our sample have the asymmetric
characteristics of one-sided jets, which is indicative of the high
degree of relativistic beaming in the sample. Five sources have
two-sided jets, of which four are low-luminosity AGN
(\S~\ref{two_sided}).  A small number of such weakly-beamed sources
are expected to be present in our flux-density limited sample, by
virtue of their proximity. We note that earlier authors have
speculated on the two-sided nature of the parsec-scale jet in the
luminous radio galaxy 2021+614
\citep{TSR00,PT00}. However, since the southernmost feature is more variable, 
more compact \citep{L01}, and has a flat spectral index \citep{BL98}
as compared to the other jet features, we consider it as the core
feature in a one-sided jet.

\subsection{\label{compact}Unresolved and Nearly Unresolved Sources}

We have found that three sources (0235+164, 1741$-$038, and 1751+288)
are effectively unresolved by the VLBA at 15 GHz at all epochs, with
essentially all of the flux density contained within a few tenths of a
milliarcsecond.  Model fitting of the visibility data for 0235+164 and
1741$-$038 indicated possible structure on a scale of $\sim 0.3$ mas;
however, we could not reliably cross identify any components across
multiple epochs for any of these sources. Another source, 1324+224, is
mostly very compact, but has a very weak feature located approximately
3 mas to the southwest of the bright core feature. However, its
measured position was erratic across multiple epochs and we were
unable to define any meaningful motions in this source. The
stacked-epoch image of 1739+522 contains a weak diffuse structure to
the north of a prominent compact core, which may be part of a highly
curved jet.

These sources could appear very compact either because their jets are
atypically weak, or they possess a highly beamed core that swamps the
jet emission in our limited sensitivity/dynamic range images. It is
also possible that they may be experiencing a period of inactivity in
which new jet components are not being ejected. We have witnessed the
latter in 1308+326, which had very little discernible jet structure
prior to 1997, and now has a pronounced mas-scale jet (Figure Set 1).

With the exception of 0235+164, none of the five compact sources
mentioned above show significant extended radio structure in deep VLA
images \citep{UJP81,PFJ82,CLK07}. They are therefore well-suited as
calibration sources for flux density and absolute polarization
position angle calibration when simultaneous single-dish/VLA and VLBA
data are available. Since any observable jet structure in these
objects must be on angular scales just at or less than the size of the
2 cm VLBA restoring beam, their kinematics would be best studied with
shorter wavelength ground and/or space VLBI, which can provide better
spatial resolution.

\subsection{\label{twosided}Two-sided Jets \label{two_sided}}

Multi-epoch observations of two-sided AGN jets are extremely useful
for testing whether apparent motions of jet features do indeed reflect
the underlying flow. In the case of random pattern speeds, we would
not necessarily expect motions in the jet and counter-jet to be correlated.
Detailed kinematic studies of two-sided AGN jets (e.g.,
\citealt*{TV97,VRK03}) have revealed similar speeds in the jet and
counter-jet, and have provided additional estimates of the viewing
angles to these jets. In this section we provide brief descriptions of
the five two-sided AGN jets in the MOJAVE sample.

0238$-$084 (NGC~1052): The morphology and kinematics of this source
have been studied extensively by \cite{VRK03}.  The jet shows numerous
features moving outward along the jet and counter-jet at speeds up to
$\sim 0.3$ c. Multi-frequency VLBA observations by \cite{VRK03} and
\cite{KRL04} indicate that the core feature is obscured at 15 GHz by strong 
free-free absorption associated with a circumnuclear torus. A more
detailed analysis of the multi-epoch structure of NGC~1052 based on
MOJAVE and additional 22 and 43 GHz data from a dedicated
multi-wavelength monitoring campaign is currently underway (E. Ros et
al., in preparation).

0316+413 (3C~84, NGC~1275): Although pre-VLBA observations indicated a
one-sided jet, albeit with complex structure, the earliest 15 GHz VLBA
observations revealed a previously undetected component which had been
obscured at the lower frequencies used previously. Extensive
observations by \cite{WDR00} have indicated that the northern lobe is
likely being affected by free-free absorption from ionized gas
associated with an accretion disk. The central feature contains
complex sub-structure and motions \citep{DKR98}, including several
distinct circularly polarized regions seen in the MOJAVE data
\citep{HL06} and at lower frequencies \citep{HW04}. 

1228+126 (M87): The 15 GHz VLBA kinematics of this nearby AGN were
reported by \cite{KLH07}, and consist of relatively slow speeds below
0.07c.  We have taken the core to be coincident with the brightest
component, for which we have also detected circular polarization
(Paper II); a faint counter-jet is observed not only at 2 cm
\citep{KLH07} but also with higher resolution at 13 and 7 mm \citep{LWJ}.

1413+135: This highly unusual two-sided BL Lac object with an apparent
spiral host galaxy has been the subject of previous VLBI studies by
\cite{PSS94,PCS96,PSC02} and \cite{WVL01}. At 15 GHz we see a mostly
one-sided jet pointing toward the west, but at longer wavelengths
there is a prominent but more diffuse jet pointing toward the
east. The line of sight to the nucleus shows HI and molecular
absorption \citep{Con99}. It is the only high-luminosity two-sided jet
in the MOJAVE sample.  Although 1413+135 has been classified as a
compact symmetric object by \cite{GPG05}, its core component is highly
variable, which is not a general characteristic of that class.

1957+405 (Cygnus A): The parsec-scale jet of this nearby radio galaxy
contains a number of well defined features in both the main jet and
the counter-jet. The jet kinematics have been studied by \cite{BKK05},
based in part on 2 cm VLBA survey and MOJAVE data, and their
multi-frequency VLBA analysis suggests that like NGC~1052 and 3C~84,
the counter-jet is partially obscured by an absorbing torus.

Although it is tempting to interpret their morphology and absorption
as evidence that these two-sided jets are plane-of-the sky versions of
the one-sided jets, \cite{CLH07} have demonstrated that the
low-luminosity two-sided sources apparently do not differ from the
high luminosity one-sided sources solely because of their
orientations. \cite{KLH07} have also argued that the prominent
parsec-scale jet in M87 may be intrinsically one-sided. The kinematic
interpretation of the two-sided jets in the MOJAVE sample will be
presented in a separate paper.

\subsection{\label{mojave2}Current monitoring program and additional sources}

The flux-density limited MOJAVE sample is statistically complete on
the basis of compact emission at 15 GHz, which is essential for
understanding the complex selection biases associated with observed
blazar samples.  However, new high energy observatories such as Fermi
are likely to detect a broader range of blazars than those derived
from a strictly radio-selected sample. For this reason, in 2006 we
expanded the MOJAVE monitoring program beyond the flux-density limited sample
to encompass 65 additional sources of interest. These include i) a
subsample of lower-luminosity AGN with 8 GHz VLBA flux density greater
than 0.35 Jy, 8 GHz VLBA luminosity $< 10^{26} \; \mathrm{W \;
Hz^{-1}}$ and $z < 0.3$ that we identified using the VLBA Calibrator
Survey \citep{BGP02,VCS2,VCS3}, ii) several additional AGN identified as
high-confidence EGRET gamma-ray associations by \cite{2003ApJ...590..109S,SRM04},
and iii) several AGN known to have particularly interesting or unusual
kinematics from our prior 2 cm VLBA survey observations
\citep{KL04}. The current AGN monitoring list can be found at the
MOJAVE website.

As of Fall 2008, the MOJAVE program is regularly monitoring 200 AGN with
the VLBA at 15 GHz, the UMRAO 26 m radio telescope at 5, 8, and 15
GHz, and the RATAN 600m telescope at 1, 2.3, 4.8, 7.7, 11, 22 GHz.  We
anticipate adding up to 100 additional AGN to our current monitoring
list, based on new gamma-ray detections made by the Fermi LAT
instrument (A. A. Abdo et al., in preparation), provided these sources
have sufficient compact flux density ($\gtrsim 100$ mJy) for imaging
with the VLBA at 15 GHz. This extended sample will ensure complete
coverage of several regions of the gamma-ray/radio flux density plane,
and will be used for correlative studies.

\section{Summary \label{summary}}

The MOJAVE program represents the largest full-polarization high
resolution monitoring program of powerful AGN jets carried out to
date. The complete flux-density limited sample of 135 sources contains the
brightest and most highly beamed AGN visible in the northern
sky. Composed primarily of blazars, it is especially useful for
comparisons with studies at other wavelengths, especially at very high
energies, where blazars tend to dominate the all-sky catalogs. In
2006, we expanded the monitoring sample to include an additional 65
AGN on the basis of their known gamma-ray emission, low luminosity,
and/or unusual jet kinematics. This list will be further supplemented
with new AGN detections from the Fermi gamma-ray observatory.

By combining archival data from the 2 cm Survey and other VLBA
programs, we have obtained 2424 individual 15 GHz VLBA images on the
flux-density limited MOJAVE sample, covering the period 1994 August to
2007 September. The full set of images is publicly available on the
MOJAVE website. These data indicate that 94\% of the jets are
one-sided, likely due to relativistic beaming effects. Of the
remainder, three AGN (0235+164, 1741$-$038, and 1751+288) are
essentially unresolved on VLBA scales, having essentially all of their
flux density contained within a few tenths of a milliarcsecond. Five
other AGN: 0238-084(NGC~1052), 0316+413 (3C~84), 1228+126 (M87),
1413+135, 1957+405 (Cygnus A), have two-sided jet morphologies. The
statistical and kinematic analysis of the full dataset will be
presented in subsequent papers in this series.

\acknowledgements 

The authors wish to acknowledge the other members of the MOJAVE team
and students who have contributed to this work, including Christian
Fromm, Kirill Sokolovsky and Alexander Pushkarev at MPIfR as well as
Andrew Merrill, Nick Mellott, Kevin O'Brien, and Amy Lankey at Purdue
University. MLL has been supported under NSF grants AST-0406923 \&
AST-0807860, and a grant from the Purdue Research Foundation. Part of
this work was done by MLL, DCH, and YYK during their Jansky
fellowships at the NRAO and also by YYK and TS during their Alexander
von Humboldt fellowships at the MPIfR. MK has been supported in
part by the NASA Postdoctoral Program at the Goddard Space Flight
Center, administered by Oak Ridge Associated Universities through a
contract with NASA. \mbox{RATAN--600} observations are partly
supported by the Russian Foundation for Basic Research (projects
01-02-16812, 05-02-17377, 08-02-00545). TS has been also supported in
part by the Max-Planck-Gesellschaft and by the Academy of Finland
grant 120516. \facility[NRAO(VLBA)]{The National Radio Astronomy
Observatory is a facility of the National Science Foundation operated
under cooperative agreement by Associated Universities, Inc.} This
research has also made use of the following resources: the University
of Michigan Radio Astronomy Observatory, which is supported by the
National Science Foundation and by funds from the University of
Michigan, NASA's Astrophysics Data System, and the NASA/IPAC
Extragalactic Database (NED). The latter is operated by the Jet
Propulsion Laboratory, California Institute of Technology, under
contract with the National Aeronautics and Space Administration.


%
%
%

\clearpage
\LongTables
\input{1gentable}
\input{2maptablestub}
\end{document}

%% file: 1gentable.tex
\begin{deluxetable*}{llrrlccll} 
\tablecolumns{9} 
\tabletypesize{\scriptsize} 
\tablewidth{0pt}  
\tablecaption{\label{gentable}The Complete Flux-Limited MOJAVE AGN Sample}  
\tablehead{\colhead{Source} & \colhead {Alias} &  
\colhead{R.A.} &\colhead{Declination} & \colhead{N} & 
\colhead{Opt. Cl.} &\colhead{Radio}  &   \colhead{z} &   \colhead{Reference for z} \\  
\colhead{(1)} & \colhead{(2)} & \colhead{(3)} & \colhead{(4)} &  
\colhead{(5)} & \colhead{(6)} & \colhead{(7)} & \colhead{(8)}  & \colhead{(9)}   } 
\startdata 
0003$-$066  & NRAO 005 & 00$\mathrm{^h}$06$\mathrm{^m}$13.8929$\mathrm{^s}$ &$-$06$\arcdeg$23$\mathrm{^\prime}$35.3353$\mathrm{^{\prime\prime}}$  & \phantom{0}19 & B & F & 0.347 &\cite{1989AAS...80..103S} \\ 
0007+106  & III Zw 2 & 00$\mathrm{^h}$10$\mathrm{^m}$31.0059$\mathrm{^s}$ &+10$\arcdeg$58$\mathrm{^\prime}$29.5044$\mathrm{^{\prime\prime}}$  & \phantom{0}15 & G & F & 0.0893 &\cite{1970ApJ...160..405S} \\ 
0016+731  & \n & 00$\mathrm{^h}$19$\mathrm{^m}$45.7864$\mathrm{^s}$ &+73$\arcdeg$27$\mathrm{^\prime}$30.0175$\mathrm{^{\prime\prime}}$  & \phantom{0}12 & Q & F & 1.781 &\cite{1996ApJS..107..541L} \\ 
0048$-$097  & \n & 00$\mathrm{^h}$50$\mathrm{^m}$41.3174$\mathrm{^s}$ &$-$09$\arcdeg$29$\mathrm{^\prime}$05.2103$\mathrm{^{\prime\prime}}$  & \phantom{0}14 & B & F & \n &\n \\ 
0059+581  & \n & 01$\mathrm{^h}$02$\mathrm{^m}$45.7624$\mathrm{^s}$ &+58$\arcdeg$24$\mathrm{^\prime}$11.1366$\mathrm{^{\prime\prime}}$  & \phantom{0}13 & Q\tablenotemark{d} & F & 0.644 &\cite{2005ApJ...626...95S} \\ 
0106+013  & \n & 01$\mathrm{^h}$08$\mathrm{^m}$38.7711$\mathrm{^s}$ &+01$\arcdeg$35$\mathrm{^\prime}$00.3173$\mathrm{^{\prime\prime}}$  & \phantom{0}15 & Q & F & 2.099 &\cite{1995AJ....109.1498H} \\ 
0109+224  & \n & 01$\mathrm{^h}$12$\mathrm{^m}$05.8247$\mathrm{^s}$ &+22$\arcdeg$44$\mathrm{^\prime}$38.7864$\mathrm{^{\prime\prime}}$  & \phantom{00}9 & B & F & 0.265 &\cite{2008ApJS..175...97H} \\ 
0119+115  & \n & 01$\mathrm{^h}$21$\mathrm{^m}$41.5950$\mathrm{^s}$ &+11$\arcdeg$49$\mathrm{^\prime}$50.4131$\mathrm{^{\prime\prime}}$  & \phantom{0}11 & Q & F & 0.570 &\cite{1994AAS..105..211S} \\ 
0133+476  & DA 55 & 01$\mathrm{^h}$36$\mathrm{^m}$58.5948$\mathrm{^s}$ &+47$\arcdeg$51$\mathrm{^\prime}$29.1001$\mathrm{^{\prime\prime}}$  & \phantom{0}25 & Q & F & 0.859 &\cite{1996ApJS..107..541L} \\ 
0202+149  & 4C +15.05 & 02$\mathrm{^h}$04$\mathrm{^m}$50.4139$\mathrm{^s}$ &+15$\arcdeg$14$\mathrm{^\prime}$11.0437$\mathrm{^{\prime\prime}}$  & \phantom{0}13 & Q\tablenotemark{c} & F & 0.405 &\cite{1998AJ....115.1253P} \\ 
0202+319  & \n & 02$\mathrm{^h}$05$\mathrm{^m}$04.9254$\mathrm{^s}$ &+32$\arcdeg$12$\mathrm{^\prime}$30.0955$\mathrm{^{\prime\prime}}$  & \phantom{0}12 & Q & F & 1.466 &\cite{1970ApJ...160L..33B} \\ 
0212+735  & \n & 02$\mathrm{^h}$17$\mathrm{^m}$30.8134$\mathrm{^s}$ &+73$\arcdeg$49$\mathrm{^\prime}$32.6218$\mathrm{^{\prime\prime}}$  & \phantom{0}14 & Q & F & 2.367 &\cite{1996ApJS..107..541L} \\ 
0215+015  & OD 026 & 02$\mathrm{^h}$17$\mathrm{^m}$48.9548$\mathrm{^s}$ &+01$\arcdeg$44$\mathrm{^\prime}$49.6991$\mathrm{^{\prime\prime}}$  & \phantom{0}17 & Q & F & 1.715 &\cite{1988AA...192....1B} \\ 
0224+671  & 4C +67.05 & 02$\mathrm{^h}$28$\mathrm{^m}$50.0515$\mathrm{^s}$ &+67$\arcdeg$21$\mathrm{^\prime}$03.0294$\mathrm{^{\prime\prime}}$  & \phantom{00}8 & Q\tablenotemark{d} & F & 0.523 &\cite{2005ApJ...626...95S} \\ 
0234+285  & CTD 20 & 02$\mathrm{^h}$37$\mathrm{^m}$52.4057$\mathrm{^s}$ &+28$\arcdeg$48$\mathrm{^\prime}$08.9901$\mathrm{^{\prime\prime}}$  & \phantom{0}16 & Q & F & 1.207 &\cite{1977ApJ...217..358S} \\ 
0235+164  & \n & 02$\mathrm{^h}$38$\mathrm{^m}$38.9301$\mathrm{^s}$ &+16$\arcdeg$36$\mathrm{^\prime}$59.2746$\mathrm{^{\prime\prime}}$  & \phantom{0}36 & B\tablenotemark{a} & F & 0.940 &\cite{1987ApJ...318..577C} \\ 
0238$-$084  & NGC 1052 & 02$\mathrm{^h}$41$\mathrm{^m}$04.7985$\mathrm{^s}$ &$-$08$\arcdeg$15$\mathrm{^\prime}$20.7518$\mathrm{^{\prime\prime}}$  & \phantom{0}30 & G & F & 0.005037 &\cite{2005MNRAS.356.1440D} \\ 
0300+470  & 4C +47.08 & 03$\mathrm{^h}$03$\mathrm{^m}$35.2422$\mathrm{^s}$ &+47$\arcdeg$16$\mathrm{^\prime}$16.2755$\mathrm{^{\prime\prime}}$  & \phantom{0}11 & B & F & \n &\n \\ 
0316+413  & 3C 84 & 03$\mathrm{^h}$19$\mathrm{^m}$48.1601$\mathrm{^s}$ &+41$\arcdeg$30$\mathrm{^\prime}$42.1048$\mathrm{^{\prime\prime}}$  & \phantom{0}22 & G & F & 0.0176 &\cite{1992ApJS...83...29S} \\ 
0333+321  & NRAO 140 & 03$\mathrm{^h}$36$\mathrm{^m}$30.1076$\mathrm{^s}$ &+32$\arcdeg$18$\mathrm{^\prime}$29.3423$\mathrm{^{\prime\prime}}$  & \phantom{0}18 & Q & F & 1.259 &\cite{1991ApJ...382..433S} \\ 
0336$-$019  & CTA 26 & 03$\mathrm{^h}$39$\mathrm{^m}$30.9378$\mathrm{^s}$ &$-$01$\arcdeg$46$\mathrm{^\prime}$35.8041$\mathrm{^{\prime\prime}}$  & \phantom{0}16 & Q & F & 0.852 &\cite{1978ApJS...36..317W} \\ 
0403$-$132  & \n & 04$\mathrm{^h}$05$\mathrm{^m}$34.0034$\mathrm{^s}$ &$-$13$\arcdeg$08$\mathrm{^\prime}$13.6907$\mathrm{^{\prime\prime}}$  & \phantom{0}10 & Q & F & 0.571 &\cite{1996ApJS..104...37M} \\ 
0415+379  & 3C 111 & 04$\mathrm{^h}$18$\mathrm{^m}$21.2772$\mathrm{^s}$ &+38$\arcdeg$01$\mathrm{^\prime}$35.8001$\mathrm{^{\prime\prime}}$  & \phantom{0}24 & G & S & 0.0491 &\cite{2004ApJS..150..181E} \\ 
0420$-$014  & \n & 04$\mathrm{^h}$23$\mathrm{^m}$15.8007$\mathrm{^s}$ &$-$01$\arcdeg$20$\mathrm{^\prime}$33.0654$\mathrm{^{\prime\prime}}$  & \phantom{0}43 & Q & F & 0.914 &\cite{1978ApJS...36..317W} \\ 
0422+004  & \n & 04$\mathrm{^h}$24$\mathrm{^m}$46.8421$\mathrm{^s}$ &+00$\arcdeg$36$\mathrm{^\prime}$06.3293$\mathrm{^{\prime\prime}}$  & \phantom{00}8 & B & F & \n &\n \\ 
0430+052  & 3C 120 & 04$\mathrm{^h}$33$\mathrm{^m}$11.0955$\mathrm{^s}$ &+05$\arcdeg$21$\mathrm{^\prime}$15.6192$\mathrm{^{\prime\prime}}$  & \phantom{0}43 & G & F & 0.033 &\cite{1988PASP..100.1423M} \\ 
0446+112  & \n & 04$\mathrm{^h}$49$\mathrm{^m}$07.6711$\mathrm{^s}$ &+11$\arcdeg$21$\mathrm{^\prime}$28.5965$\mathrm{^{\prime\prime}}$  & \phantom{00}9 & U\tablenotemark{b} & F & \n &\n \\ 
0458$-$020  & \n & 05$\mathrm{^h}$01$\mathrm{^m}$12.8099$\mathrm{^s}$ &$-$01$\arcdeg$59$\mathrm{^\prime}$14.2562$\mathrm{^{\prime\prime}}$  & \phantom{0}10 & Q & F & 2.286 &\cite{1974ApJ...190..509S} \\ 
0528+134  & \n & 05$\mathrm{^h}$30$\mathrm{^m}$56.4167$\mathrm{^s}$ &+13$\arcdeg$31$\mathrm{^\prime}$55.1496$\mathrm{^{\prime\prime}}$  & \phantom{0}24 & Q & F & 2.070 &\cite{1993ApJ...409..134H} \\ 
0529+075  & \n & 05$\mathrm{^h}$32$\mathrm{^m}$38.9985$\mathrm{^s}$ &+07$\arcdeg$32$\mathrm{^\prime}$43.3451$\mathrm{^{\prime\prime}}$  & \phantom{00}6 & Q\tablenotemark{d} & F & 1.254 &\cite{2005ApJ...626...95S} \\ 
0529+483  & \n & 05$\mathrm{^h}$33$\mathrm{^m}$15.8658$\mathrm{^s}$ &+48$\arcdeg$22$\mathrm{^\prime}$52.8079$\mathrm{^{\prime\prime}}$  & \phantom{0}10 & Q & F & 1.162 &\cite{2003AJ....125..572H} \\ 
0552+398  & DA 193 & 05$\mathrm{^h}$55$\mathrm{^m}$30.8056$\mathrm{^s}$ &+39$\arcdeg$48$\mathrm{^\prime}$49.1650$\mathrm{^{\prime\prime}}$  & \phantom{0}28 & Q & F & 2.363 &\cite{1999ApJ...514...40M} \\ 
0605$-$085  & OC $-$010 & 06$\mathrm{^h}$07$\mathrm{^m}$59.6992$\mathrm{^s}$ &$-$08$\arcdeg$34$\mathrm{^\prime}$49.9782$\mathrm{^{\prime\prime}}$  & \phantom{0}15 & Q & F & 0.872 &\cite{1993AAS...97..483S} \\ 
0607$-$157  & \n & 06$\mathrm{^h}$09$\mathrm{^m}$40.9495$\mathrm{^s}$ &$-$15$\arcdeg$42$\mathrm{^\prime}$40.6726$\mathrm{^{\prime\prime}}$  & \phantom{0}20 & Q & F & 0.324 &\cite{1978MNRAS.185..149H} \\ 
0642+449  & OH 471 & 06$\mathrm{^h}$46$\mathrm{^m}$32.0260$\mathrm{^s}$ &+44$\arcdeg$51$\mathrm{^\prime}$16.5901$\mathrm{^{\prime\prime}}$  & \phantom{0}10 & Q & F & 3.396 &\cite{1994ApJ...436..678O} \\ 
0648$-$165  & \n & 06$\mathrm{^h}$50$\mathrm{^m}$24.5819$\mathrm{^s}$ &$-$16$\arcdeg$37$\mathrm{^\prime}$39.7253$\mathrm{^{\prime\prime}}$  & \phantom{00}6 & U\tablenotemark{d} & F & \n &\n \\ 
0716+714  & \n & 07$\mathrm{^h}$21$\mathrm{^m}$53.4485$\mathrm{^s}$ &+71$\arcdeg$20$\mathrm{^\prime}$36.3634$\mathrm{^{\prime\prime}}$  & \phantom{0}26 & B & F & 0.310 &\cite{2008AA...487L..29N} \\ 
0727$-$115  & \n & 07$\mathrm{^h}$30$\mathrm{^m}$19.1125$\mathrm{^s}$ &$-$11$\arcdeg$41$\mathrm{^\prime}$12.6005$\mathrm{^{\prime\prime}}$  & \phantom{0}11 & Q & F & 1.591 &\cite{2002AJ....124..662Z} \\ 
0730+504  & \n & 07$\mathrm{^h}$33$\mathrm{^m}$52.5206$\mathrm{^s}$ &+50$\arcdeg$22$\mathrm{^\prime}$09.0621$\mathrm{^{\prime\prime}}$  & \phantom{00}7 & Q & F & 0.720 &\cite{1997MNRAS.290..380H} \\ 
0735+178  & OI 158 & 07$\mathrm{^h}$38$\mathrm{^m}$07.3937$\mathrm{^s}$ &+17$\arcdeg$42$\mathrm{^\prime}$18.9982$\mathrm{^{\prime\prime}}$  & \phantom{0}24 & B & F & \n &\n \\ 
0736+017  & OI 061 & 07$\mathrm{^h}$39$\mathrm{^m}$18.0339$\mathrm{^s}$ &+01$\arcdeg$37$\mathrm{^\prime}$04.6178$\mathrm{^{\prime\prime}}$  & \phantom{0}14 & Q & F & 0.191 &\cite{1967ApJ...147..837L} \\ 
0738+313  & OI 363 & 07$\mathrm{^h}$41$\mathrm{^m}$10.7033$\mathrm{^s}$ &+31$\arcdeg$12$\mathrm{^\prime}$00.2291$\mathrm{^{\prime\prime}}$  & \phantom{0}17 & Q & F & 0.631 &\cite{SDSS2} \\ 
0742+103  & \n & 07$\mathrm{^h}$45$\mathrm{^m}$33.0595$\mathrm{^s}$ &+10$\arcdeg$11$\mathrm{^\prime}$12.6923$\mathrm{^{\prime\prime}}$  & \phantom{0}15 & Q\tablenotemark{d} & P & 2.624 &\cite{2003MNRAS.346.1021B} \\ 
0748+126  & \n & 07$\mathrm{^h}$50$\mathrm{^m}$52.0457$\mathrm{^s}$ &+12$\arcdeg$31$\mathrm{^\prime}$04.8282$\mathrm{^{\prime\prime}}$  & \phantom{0}16 & Q & F & 0.889 &\cite{1979ApJ...232..400P} \\ 
0754+100  & \n & 07$\mathrm{^h}$57$\mathrm{^m}$06.6429$\mathrm{^s}$ &+09$\arcdeg$56$\mathrm{^\prime}$34.8525$\mathrm{^{\prime\prime}}$  & \phantom{0}16 & B & F & 0.266 &\cite{2003AA...412..651C} \\ 
0804+499  & \n & 08$\mathrm{^h}$08$\mathrm{^m}$39.6663$\mathrm{^s}$ &+49$\arcdeg$50$\mathrm{^\prime}$36.5304$\mathrm{^{\prime\prime}}$  & \phantom{0}12 & Q & F & 1.436 &\cite{SDSS4} \\ 
0805$-$077  & \n & 08$\mathrm{^h}$08$\mathrm{^m}$15.5360$\mathrm{^s}$ &$-$07$\arcdeg$51$\mathrm{^\prime}$09.8864$\mathrm{^{\prime\prime}}$  & \phantom{00}7 & Q & F & 1.837 &\cite{1988ApJ...327..561W} \\ 
0808+019  & \n & 08$\mathrm{^h}$11$\mathrm{^m}$26.7073$\mathrm{^s}$ &+01$\arcdeg$46$\mathrm{^\prime}$52.2202$\mathrm{^{\prime\prime}}$  & \phantom{0}11 & B & F & 1.148 &\cite{2005AJ....129..559S} \\ 
0814+425  & OJ 425 & 08$\mathrm{^h}$18$\mathrm{^m}$15.9996$\mathrm{^s}$ &+42$\arcdeg$22$\mathrm{^\prime}$45.4149$\mathrm{^{\prime\prime}}$  & \phantom{0}15 & B & F & 0.245 &\cite{2008AA...484..119B} \\ 
0823+033  & \n & 08$\mathrm{^h}$25$\mathrm{^m}$50.3384$\mathrm{^s}$ &+03$\arcdeg$09$\mathrm{^\prime}$24.5201$\mathrm{^{\prime\prime}}$  & \phantom{0}22 & B & F & 0.506 &\cite{1993AAS...98..393S} \\ 
0827+243  & OJ 248 & 08$\mathrm{^h}$30$\mathrm{^m}$52.0862$\mathrm{^s}$ &+24$\arcdeg$10$\mathrm{^\prime}$59.8204$\mathrm{^{\prime\prime}}$  & \phantom{0}19 & Q & F & 0.940 &\cite{SDSS5} \\ 
0829+046  & OJ 049 & 08$\mathrm{^h}$31$\mathrm{^m}$48.8770$\mathrm{^s}$ &+04$\arcdeg$29$\mathrm{^\prime}$39.0859$\mathrm{^{\prime\prime}}$  & \phantom{0}26 & B & F & 0.174 &\cite{SDSS3} \\ 
0836+710  & 4C 71.07 & 08$\mathrm{^h}$41$\mathrm{^m}$24.3653$\mathrm{^s}$ &+70$\arcdeg$53$\mathrm{^\prime}$42.1730$\mathrm{^{\prime\prime}}$  & \phantom{0}13 & Q & F & 2.218 &\cite{1999ApJ...514...40M} \\ 
0838+133  & 3C 207 & 08$\mathrm{^h}$40$\mathrm{^m}$47.5884$\mathrm{^s}$ &+13$\arcdeg$12$\mathrm{^\prime}$23.5641$\mathrm{^{\prime\prime}}$  & \phantom{0}16 & Q & F & 0.681 &\cite{1996ApJS..104...37M} \\ 
0851+202  & OJ 287 & 08$\mathrm{^h}$54$\mathrm{^m}$48.8749$\mathrm{^s}$ &+20$\arcdeg$06$\mathrm{^\prime}$30.6409$\mathrm{^{\prime\prime}}$  & \phantom{0}69 & B & F & 0.306 &\cite{1989AAS...80..103S} \\ 
0906+015  & 4C +01.24 & 09$\mathrm{^h}$09$\mathrm{^m}$10.0916$\mathrm{^s}$ &+01$\arcdeg$21$\mathrm{^\prime}$35.6180$\mathrm{^{\prime\prime}}$  & \phantom{0}12 & Q & F & 1.024 &\cite{SDSS2} \\ 
0917+624  & \n & 09$\mathrm{^h}$21$\mathrm{^m}$36.2311$\mathrm{^s}$ &+62$\arcdeg$15$\mathrm{^\prime}$52.1804$\mathrm{^{\prime\prime}}$  & \phantom{0}13 & Q & F & 1.446 &\cite{1993AAS..101..521S} \\ 
0923+392  & 4C +39.25 & 09$\mathrm{^h}$27$\mathrm{^m}$03.0139$\mathrm{^s}$ &+39$\arcdeg$02$\mathrm{^\prime}$20.8519$\mathrm{^{\prime\prime}}$  & \phantom{0}26 & Q & F & 0.695 &\cite{SDSS3} \\ 
0945+408  & 4C +40.24 & 09$\mathrm{^h}$48$\mathrm{^m}$55.3382$\mathrm{^s}$ &+40$\arcdeg$39$\mathrm{^\prime}$44.5869$\mathrm{^{\prime\prime}}$  & \phantom{0}13 & Q & F & 1.249 &\cite{SDSS3} \\ 
0955+476  & \n & 09$\mathrm{^h}$58$\mathrm{^m}$19.6716$\mathrm{^s}$ &+47$\arcdeg$25$\mathrm{^\prime}$07.8424$\mathrm{^{\prime\prime}}$  & \phantom{00}7 & Q & F & 1.882 &\cite{SDSS2} \\ 
1036+054  & \n & 10$\mathrm{^h}$38$\mathrm{^m}$46.7799$\mathrm{^s}$ &+05$\arcdeg$12$\mathrm{^\prime}$29.0867$\mathrm{^{\prime\prime}}$  & \phantom{00}7 & Q\tablenotemark{d} & F & 0.473 &\cite{2008ApJS..175...97H} \\ 
1038+064  & 4C +06.41 & 10$\mathrm{^h}$41$\mathrm{^m}$17.1625$\mathrm{^s}$ &+06$\arcdeg$10$\mathrm{^\prime}$16.9235$\mathrm{^{\prime\prime}}$  & \phantom{00}8 & Q & F & 1.265 &\cite{1991ApJ...382..433S} \\ 
1045$-$188  & \n & 10$\mathrm{^h}$48$\mathrm{^m}$06.6206$\mathrm{^s}$ &$-$19$\arcdeg$09$\mathrm{^\prime}$35.7268$\mathrm{^{\prime\prime}}$  & \phantom{00}6 & Q & F & 0.595 &\cite{1993AAS...97..483S} \\ 
1055+018  & 4C +01.28 & 10$\mathrm{^h}$58$\mathrm{^m}$29.6052$\mathrm{^s}$ &+01$\arcdeg$33$\mathrm{^\prime}$58.8237$\mathrm{^{\prime\prime}}$  & \phantom{0}27 & Q & F & 0.890 &\cite{2004MNRAS.349.1397C} \\ 
1124$-$186  & \n & 11$\mathrm{^h}$27$\mathrm{^m}$04.3925$\mathrm{^s}$ &$-$18$\arcdeg$57$\mathrm{^\prime}$17.4417$\mathrm{^{\prime\prime}}$  & \phantom{00}7 & Q & F & 1.048 &\cite{1997MNRAS.284...85D} \\ 
1127$-$145  & \n & 11$\mathrm{^h}$30$\mathrm{^m}$07.0526$\mathrm{^s}$ &$-$14$\arcdeg$49$\mathrm{^\prime}$27.3882$\mathrm{^{\prime\prime}}$  & \phantom{0}16 & Q & F & 1.184 &\cite{1986MNRAS.218..331W} \\ 
1150+812  & \n & 11$\mathrm{^h}$53$\mathrm{^m}$12.4992$\mathrm{^s}$ &+80$\arcdeg$58$\mathrm{^\prime}$29.1546$\mathrm{^{\prime\prime}}$  & \phantom{0}11 & Q & F & 1.250 &\cite{1986AA...168...17E} \\ 
1156+295  & 4C +29.45 & 11$\mathrm{^h}$59$\mathrm{^m}$31.8339$\mathrm{^s}$ &+29$\arcdeg$14$\mathrm{^\prime}$43.8269$\mathrm{^{\prime\prime}}$  & \phantom{0}30 & Q & F & 0.729 &\cite{1994ApJS...93....1A} \\ 
1213$-$172  & \n & 12$\mathrm{^h}$15$\mathrm{^m}$46.7518$\mathrm{^s}$ &$-$17$\arcdeg$31$\mathrm{^\prime}$45.4029$\mathrm{^{\prime\prime}}$  & \phantom{00}7 & U\tablenotemark{d} & F & \n &\n \\ 
1219+044  & \n & 12$\mathrm{^h}$22$\mathrm{^m}$22.5496$\mathrm{^s}$ &+04$\arcdeg$13$\mathrm{^\prime}$15.7761$\mathrm{^{\prime\prime}}$  & \phantom{00}6 & Q & F & 0.965 &\cite{1986MNRAS.218..331W} \\ 
1222+216  & \n & 12$\mathrm{^h}$24$\mathrm{^m}$54.4584$\mathrm{^s}$ &+21$\arcdeg$22$\mathrm{^\prime}$46.3887$\mathrm{^{\prime\prime}}$  & \phantom{0}16 & Q & F & 0.432 &\cite{1987ApJ...323..108O} \\ 
1226+023  & 3C 273 & 12$\mathrm{^h}$29$\mathrm{^m}$06.6997$\mathrm{^s}$ &+02$\arcdeg$03$\mathrm{^\prime}$08.5981$\mathrm{^{\prime\prime}}$  & \phantom{0}54 & Q & F & 0.158 &\cite{1992ApJS...83...29S} \\ 
1228+126  & M87 & 12$\mathrm{^h}$30$\mathrm{^m}$49.4234$\mathrm{^s}$ &+12$\arcdeg$23$\mathrm{^\prime}$28.0438$\mathrm{^{\prime\prime}}$  & \phantom{0}25 & G & S & 0.00436 &\cite{2000MNRAS.313..469S} \\ 
1253$-$055  & 3C 279 & 12$\mathrm{^h}$56$\mathrm{^m}$11.1666$\mathrm{^s}$ &$-$05$\arcdeg$47$\mathrm{^\prime}$21.5247$\mathrm{^{\prime\prime}}$  & \phantom{0}89 & Q & F & 0.536 &\cite{1996ApJS..104...37M} \\ 
1308+326  & \n & 13$\mathrm{^h}$10$\mathrm{^m}$28.6639$\mathrm{^s}$ &+32$\arcdeg$20$\mathrm{^\prime}$43.7829$\mathrm{^{\prime\prime}}$  & \phantom{0}44 & Q & F & 0.997 &\cite{1992ApJS...79....1T} \\ 
1324+224  & \n & 13$\mathrm{^h}$27$\mathrm{^m}$00.8613$\mathrm{^s}$ &+22$\arcdeg$10$\mathrm{^\prime}$50.1630$\mathrm{^{\prime\prime}}$  & \phantom{00}8 & Q & F & 1.400 &\cite{1996MNRAS.282.1274H} \\ 
1334$-$127  & \n & 13$\mathrm{^h}$37$\mathrm{^m}$39.7828$\mathrm{^s}$ &$-$12$\arcdeg$57$\mathrm{^\prime}$24.6933$\mathrm{^{\prime\prime}}$  & \phantom{0}16 & Q & F & 0.539 &\cite{1993AAS...97..483S} \\ 
1413+135  & \n & 14$\mathrm{^h}$15$\mathrm{^m}$58.8175$\mathrm{^s}$ &+13$\arcdeg$20$\mathrm{^\prime}$23.7128$\mathrm{^{\prime\prime}}$  & \phantom{0}18 & B & F & 0.247 &\cite{1997AA...328...48W} \\ 
1417+385  & \n & 14$\mathrm{^h}$19$\mathrm{^m}$46.6138$\mathrm{^s}$ &+38$\arcdeg$21$\mathrm{^\prime}$48.4751$\mathrm{^{\prime\prime}}$  & \phantom{00}6 & Q & F & 1.831 &\cite{SDSS4} \\ 
1458+718  & 3C 309.1 & 14$\mathrm{^h}$59$\mathrm{^m}$07.5839$\mathrm{^s}$ &+71$\arcdeg$40$\mathrm{^\prime}$19.8665$\mathrm{^{\prime\prime}}$  & \phantom{00}9 & Q & S\tablenotemark{e} & 0.904 &\cite{1989ESOSR...7....1V} \\ 
1502+106  & 4C +10.39 & 15$\mathrm{^h}$04$\mathrm{^m}$24.9798$\mathrm{^s}$ &+10$\arcdeg$29$\mathrm{^\prime}$39.1986$\mathrm{^{\prime\prime}}$  & \phantom{0}11 & Q & F & 1.839 &\cite{2003ApJS..148..275A} \\ 
1504$-$166  & \n & 15$\mathrm{^h}$07$\mathrm{^m}$04.7870$\mathrm{^s}$ &$-$16$\arcdeg$52$\mathrm{^\prime}$30.2670$\mathrm{^{\prime\prime}}$  & \phantom{00}7 & Q & F & 0.876 &\cite{1978MNRAS.185..149H} \\ 
1510$-$089  & \n & 15$\mathrm{^h}$12$\mathrm{^m}$50.5329$\mathrm{^s}$ &$-$09$\arcdeg$05$\mathrm{^\prime}$59.8297$\mathrm{^{\prime\prime}}$  & \phantom{0}31 & Q & F & 0.360 &\cite{1990PASP..102.1235T} \\ 
1538+149  & 4C +14.60 & 15$\mathrm{^h}$40$\mathrm{^m}$49.4915$\mathrm{^s}$ &+14$\arcdeg$47$\mathrm{^\prime}$45.8848$\mathrm{^{\prime\prime}}$  & \phantom{0}13 & B\tablenotemark{a} & F & 0.605 &\cite{1993AAS...98..393S} \\ 
1546+027  & \n & 15$\mathrm{^h}$49$\mathrm{^m}$29.4368$\mathrm{^s}$ &+02$\arcdeg$37$\mathrm{^\prime}$01.1634$\mathrm{^{\prime\prime}}$  & \phantom{0}15 & Q & F & 0.414 &\cite{SDSS2} \\ 
1548+056  & 4C +05.64 & 15$\mathrm{^h}$50$\mathrm{^m}$35.2692$\mathrm{^s}$ &+05$\arcdeg$27$\mathrm{^\prime}$10.4484$\mathrm{^{\prime\prime}}$  & \phantom{0}13 & Q & F & 1.422 &\cite{1988ApJ...327..561W} \\ 
1606+106  & 4C +10.45 & 16$\mathrm{^h}$08$\mathrm{^m}$46.2032$\mathrm{^s}$ &+10$\arcdeg$29$\mathrm{^\prime}$07.7758$\mathrm{^{\prime\prime}}$  & \phantom{0}13 & Q & F & 1.226 &\cite{1994AAS..105...67S} \\ 
1611+343  & DA 406 & 16$\mathrm{^h}$13$\mathrm{^m}$41.0642$\mathrm{^s}$ &+34$\arcdeg$12$\mathrm{^\prime}$47.9089$\mathrm{^{\prime\prime}}$  & \phantom{0}23 & Q & F & 1.397 &\cite{SDSS5} \\ 
1633+382  & 4C +38.41 & 16$\mathrm{^h}$35$\mathrm{^m}$15.4930$\mathrm{^s}$ &+38$\arcdeg$08$\mathrm{^\prime}$04.5006$\mathrm{^{\prime\prime}}$  & \phantom{0}28 & Q & F & 1.814 &\cite{SDSS3} \\ 
1637+574  & OS 562 & 16$\mathrm{^h}$38$\mathrm{^m}$13.4563$\mathrm{^s}$ &+57$\arcdeg$20$\mathrm{^\prime}$23.9790$\mathrm{^{\prime\prime}}$  & \phantom{0}13 & Q & F & 0.751 &\cite{1996ApJS..104...37M} \\ 
1638+398  & NRAO 512 & 16$\mathrm{^h}$40$\mathrm{^m}$29.6328$\mathrm{^s}$ &+39$\arcdeg$46$\mathrm{^\prime}$46.0285$\mathrm{^{\prime\prime}}$  & \phantom{0}16 & Q & F & 1.666 &\cite{1989AAS...80..103S} \\ 
1641+399  & 3C 345 & 16$\mathrm{^h}$42$\mathrm{^m}$58.8100$\mathrm{^s}$ &+39$\arcdeg$48$\mathrm{^\prime}$36.9940$\mathrm{^{\prime\prime}}$  & \phantom{0}39 & Q & F & 0.593 &\cite{1996ApJS..104...37M} \\ 
1655+077  & \n & 16$\mathrm{^h}$58$\mathrm{^m}$09.0115$\mathrm{^s}$ &+07$\arcdeg$41$\mathrm{^\prime}$27.5405$\mathrm{^{\prime\prime}}$  & \phantom{0}14 & Q & F & 0.621 &\cite{1986MNRAS.218..331W} \\ 
1726+455  & \n & 17$\mathrm{^h}$27$\mathrm{^m}$27.6508$\mathrm{^s}$ &+45$\arcdeg$30$\mathrm{^\prime}$39.7313$\mathrm{^{\prime\prime}}$  & \phantom{00}5 & Q & F & 0.717 &\cite{1997MNRAS.290..380H} \\ 
1730$-$130  & NRAO 530 & 17$\mathrm{^h}$33$\mathrm{^m}$02.7058$\mathrm{^s}$ &$-$13$\arcdeg$04$\mathrm{^\prime}$49.5483$\mathrm{^{\prime\prime}}$  & \phantom{0}16 & Q & F & 0.902 &\cite{1984PASP...96..539J} \\ 
1739+522  & 4C +51.37 & 17$\mathrm{^h}$40$\mathrm{^m}$36.9778$\mathrm{^s}$ &+52$\arcdeg$11$\mathrm{^\prime}$43.4074$\mathrm{^{\prime\prime}}$  & \phantom{0}19 & Q & F & 1.379 &\cite{1984MNRAS.211..105W} \\ 
1741$-$038  & \n & 17$\mathrm{^h}$43$\mathrm{^m}$58.8561$\mathrm{^s}$ &$-$03$\arcdeg$50$\mathrm{^\prime}$04.6167$\mathrm{^{\prime\prime}}$  & \phantom{0}12 & Q & F & 1.054 &\cite{1988ApJ...327..561W} \\ 
1749+096  & OT 081 & 17$\mathrm{^h}$51$\mathrm{^m}$32.8186$\mathrm{^s}$ &+09$\arcdeg$39$\mathrm{^\prime}$00.7284$\mathrm{^{\prime\prime}}$  & \phantom{0}53 & B\tablenotemark{a} & F & 0.322 &\cite{1988AA...191L..16S} \\ 
1751+288  & \n & 17$\mathrm{^h}$53$\mathrm{^m}$42.4736$\mathrm{^s}$ &+28$\arcdeg$48$\mathrm{^\prime}$04.9388$\mathrm{^{\prime\prime}}$  & \phantom{00}6 & Q\tablenotemark{d} & F & 1.118 &\cite{2008ApJS..175...97H} \\ 
1758+388  & \n & 18$\mathrm{^h}$00$\mathrm{^m}$24.7654$\mathrm{^s}$ &+38$\arcdeg$48$\mathrm{^\prime}$30.6975$\mathrm{^{\prime\prime}}$  & \phantom{0}11 & Q & F & 2.092 &\cite{1994AAS..103..349S} \\ 
1800+440  & \n & 18$\mathrm{^h}$01$\mathrm{^m}$32.3148$\mathrm{^s}$ &+44$\arcdeg$04$\mathrm{^\prime}$21.9002$\mathrm{^{\prime\prime}}$  & \phantom{0}12 & Q & F & 0.663 &\cite{1982MNRAS.200..191W} \\ 
1803+784  & \n & 18$\mathrm{^h}$00$\mathrm{^m}$45.6839$\mathrm{^s}$ &+78$\arcdeg$28$\mathrm{^\prime}$04.0184$\mathrm{^{\prime\prime}}$  & \phantom{0}32 & B\tablenotemark{a} & F & 0.680 &\cite{2001AJ....122..565R} \\ 
1807+698  & 3C 371 & 18$\mathrm{^h}$06$\mathrm{^m}$50.6806$\mathrm{^s}$ &+69$\arcdeg$49$\mathrm{^\prime}$28.1085$\mathrm{^{\prime\prime}}$  & \phantom{0}21 & B & F & 0.051 &\cite{1992AAS...96..389d} \\ 
1823+568  & 4C +56.27 & 18$\mathrm{^h}$24$\mathrm{^m}$07.0684$\mathrm{^s}$ &+56$\arcdeg$51$\mathrm{^\prime}$01.4908$\mathrm{^{\prime\prime}}$  & \phantom{0}41 & B\tablenotemark{a} & F & 0.664 &\cite{1986AJ.....91..494L} \\ 
1828+487  & 3C 380 & 18$\mathrm{^h}$29$\mathrm{^m}$31.7809$\mathrm{^s}$ &+48$\arcdeg$44$\mathrm{^\prime}$46.1608$\mathrm{^{\prime\prime}}$  & \phantom{0}12 & Q & S\tablenotemark{e} & 0.692 &\cite{1996ApJS..107..541L} \\ 
1849+670  & \n & 18$\mathrm{^h}$49$\mathrm{^m}$16.0723$\mathrm{^s}$ &+67$\arcdeg$05$\mathrm{^\prime}$41.6803$\mathrm{^{\prime\prime}}$  & \phantom{00}8 & Q & F & 0.657 &\cite{1993AAS..100..395S} \\ 
1928+738  & 4C +73.18 & 19$\mathrm{^h}$27$\mathrm{^m}$48.4952$\mathrm{^s}$ &+73$\arcdeg$58$\mathrm{^\prime}$01.5698$\mathrm{^{\prime\prime}}$  & \phantom{0}33 & Q & F & 0.302 &\cite{1996ApJS..104...37M} \\ 
1936$-$155  & \n & 19$\mathrm{^h}$39$\mathrm{^m}$26.6577$\mathrm{^s}$ &$-$15$\arcdeg$25$\mathrm{^\prime}$43.0585$\mathrm{^{\prime\prime}}$  & \phantom{00}8 & Q & F & 1.657 &\cite{1984ApJ...286..498J} \\ 
1957+405  & Cygnus A & 19$\mathrm{^h}$59$\mathrm{^m}$28.3565$\mathrm{^s}$ &+40$\arcdeg$44$\mathrm{^\prime}$02.0968$\mathrm{^{\prime\prime}}$  & \phantom{0}21 & G & S & 0.0561 &\cite{1997ApJ...488L..15O} \\ 
1958$-$179  & \n & 20$\mathrm{^h}$00$\mathrm{^m}$57.0904$\mathrm{^s}$ &$-$17$\arcdeg$48$\mathrm{^\prime}$57.6726$\mathrm{^{\prime\prime}}$  & \phantom{00}8 & Q & F & 0.650 &\cite{1975MNRAS.173p..87B} \\ 
2005+403  & \n & 20$\mathrm{^h}$07$\mathrm{^m}$44.9448$\mathrm{^s}$ &+40$\arcdeg$29$\mathrm{^\prime}$48.6041$\mathrm{^{\prime\prime}}$  & \phantom{0}19 & Q & F & 1.736 &\cite{1976MNRAS.177p..43B} \\ 
2008$-$159  & \n & 20$\mathrm{^h}$11$\mathrm{^m}$15.7109$\mathrm{^s}$ &$-$15$\arcdeg$46$\mathrm{^\prime}$40.2537$\mathrm{^{\prime\prime}}$  & \phantom{00}7 & Q & F & 1.180 &\cite{1979ApJ...232..400P} \\ 
2021+317  & 4C +31.56 & 20$\mathrm{^h}$23$\mathrm{^m}$19.0173$\mathrm{^s}$ &+31$\arcdeg$53$\mathrm{^\prime}$02.3061$\mathrm{^{\prime\prime}}$  & \phantom{0}12 & U\tablenotemark{d} & F & \n &\n \\ 
2021+614  & OW 637 & 20$\mathrm{^h}$22$\mathrm{^m}$06.6818$\mathrm{^s}$ &+61$\arcdeg$36$\mathrm{^\prime}$58.8050$\mathrm{^{\prime\prime}}$  & \phantom{0}18 & G & F & 0.227 &\cite{1991ApJS...75..297H} \\ 
2037+511  & 3C 418 & 20$\mathrm{^h}$38$\mathrm{^m}$37.0347$\mathrm{^s}$ &+51$\arcdeg$19$\mathrm{^\prime}$12.6626$\mathrm{^{\prime\prime}}$  & \phantom{00}5 & Q & F & 1.686 &\cite{1985PASP...97..932S} \\ 
2121+053  & \n & 21$\mathrm{^h}$23$\mathrm{^m}$44.5174$\mathrm{^s}$ &+05$\arcdeg$35$\mathrm{^\prime}$22.0931$\mathrm{^{\prime\prime}}$  & \phantom{0}10 & Q & F & 1.941 &\cite{1991ApJ...382..433S} \\ 
2128$-$123  & \n & 21$\mathrm{^h}$31$\mathrm{^m}$35.2618$\mathrm{^s}$ &$-$12$\arcdeg$07$\mathrm{^\prime}$04.7960$\mathrm{^{\prime\prime}}$  & \phantom{0}10 & Q & F & 0.501 &\cite{1968ApJ...154L.101S} \\ 
2131$-$021  & 4C $-$02.81 & 21$\mathrm{^h}$34$\mathrm{^m}$10.3096$\mathrm{^s}$ &$-$01$\arcdeg$53$\mathrm{^\prime}$17.2387$\mathrm{^{\prime\prime}}$  & \phantom{0}17 & B\tablenotemark{a} & F & 1.285 &\cite{1997MNRAS.284...85D} \\ 
2134+004  & \n & 21$\mathrm{^h}$36$\mathrm{^m}$38.5863$\mathrm{^s}$ &+00$\arcdeg$41$\mathrm{^\prime}$54.2128$\mathrm{^{\prime\prime}}$  & \phantom{0}30 & Q & F & 1.932 &\cite{1994ApJ...436..678O} \\ 
2136+141  & OX 161 & 21$\mathrm{^h}$39$\mathrm{^m}$01.3093$\mathrm{^s}$ &+14$\arcdeg$23$\mathrm{^\prime}$35.9922$\mathrm{^{\prime\prime}}$  & \phantom{0}14 & Q & F & 2.427 &\cite{1974ApJ...190..271W} \\ 
2145+067  & 4C +06.69 & 21$\mathrm{^h}$48$\mathrm{^m}$05.4587$\mathrm{^s}$ &+06$\arcdeg$57$\mathrm{^\prime}$38.6042$\mathrm{^{\prime\prime}}$  & \phantom{0}17 & Q & F & 0.990 &\cite{1991ApJ...382..433S} \\ 
2155$-$152  & \n & 21$\mathrm{^h}$58$\mathrm{^m}$06.2819$\mathrm{^s}$ &$-$15$\arcdeg$01$\mathrm{^\prime}$09.3277$\mathrm{^{\prime\prime}}$  & \phantom{0}12 & Q & F & 0.672 &\cite{1988ApJ...327..561W} \\ 
2200+420  & BL Lac & 22$\mathrm{^h}$02$\mathrm{^m}$43.2914$\mathrm{^s}$ &+42$\arcdeg$16$\mathrm{^\prime}$39.9799$\mathrm{^{\prime\prime}}$  & \phantom{0}63 & B & F & 0.0686 &\cite{1995ApJ...452L...5V} \\ 
2201+171  & \n & 22$\mathrm{^h}$03$\mathrm{^m}$26.8937$\mathrm{^s}$ &+17$\arcdeg$25$\mathrm{^\prime}$48.2477$\mathrm{^{\prime\prime}}$  & \phantom{00}7 & Q & F & 1.076 &\cite{1977ApJ...215..427S} \\ 
2201+315  & 4C +31.63 & 22$\mathrm{^h}$03$\mathrm{^m}$14.9758$\mathrm{^s}$ &+31$\arcdeg$45$\mathrm{^\prime}$38.2700$\mathrm{^{\prime\prime}}$  & \phantom{0}15 & Q & F & 0.295 &\cite{1996ApJS..104...37M} \\ 
2209+236  & \n & 22$\mathrm{^h}$12$\mathrm{^m}$05.9663$\mathrm{^s}$ &+23$\arcdeg$55$\mathrm{^\prime}$40.5439$\mathrm{^{\prime\prime}}$  & \phantom{0}11 & Q & F & 1.125 &\cite{2003ApJ...590..109S} \\ 
2216$-$038  & \n & 22$\mathrm{^h}$18$\mathrm{^m}$52.0377$\mathrm{^s}$ &$-$03$\arcdeg$35$\mathrm{^\prime}$36.8795$\mathrm{^{\prime\prime}}$  & \phantom{00}8 & Q & F & 0.901 &\cite{1967ApJ...147..837L} \\ 
2223$-$052  & 3C 446 & 22$\mathrm{^h}$25$\mathrm{^m}$47.2593$\mathrm{^s}$ &$-$04$\arcdeg$57$\mathrm{^\prime}$01.3908$\mathrm{^{\prime\prime}}$  & \phantom{0}23 & Q & F & 1.404 &\cite{1983MNRAS.205..793W} \\ 
2227$-$088  & PHL 5225 & 22$\mathrm{^h}$29$\mathrm{^m}$40.0843$\mathrm{^s}$ &$-$08$\arcdeg$32$\mathrm{^\prime}$54.4354$\mathrm{^{\prime\prime}}$  & \phantom{00}9 & Q & F & 1.560 &\cite{SDSS2} \\ 
2230+114  & CTA 102 & 22$\mathrm{^h}$32$\mathrm{^m}$36.4089$\mathrm{^s}$ &+11$\arcdeg$43$\mathrm{^\prime}$50.9041$\mathrm{^{\prime\prime}}$  & \phantom{0}24 & Q & F & 1.037 &\cite{1994ApJS...93..125F} \\ 
2243$-$123  & \n & 22$\mathrm{^h}$46$\mathrm{^m}$18.2320$\mathrm{^s}$ &$-$12$\arcdeg$06$\mathrm{^\prime}$51.2775$\mathrm{^{\prime\prime}}$  & \phantom{0}11 & Q & F & 0.632 &\cite{1975MNRAS.173p..87B} \\ 
2251+158  & 3C 454.3 & 22$\mathrm{^h}$53$\mathrm{^m}$57.7479$\mathrm{^s}$ &+16$\arcdeg$08$\mathrm{^\prime}$53.5609$\mathrm{^{\prime\prime}}$  & \phantom{0}53 & Q & F & 0.859 &\cite{1991MNRAS.250..414J} \\ 
2331+073  & \n & 23$\mathrm{^h}$34$\mathrm{^m}$12.8282$\mathrm{^s}$ &+07$\arcdeg$36$\mathrm{^\prime}$27.5510$\mathrm{^{\prime\prime}}$  & \phantom{00}6 & Q\tablenotemark{d} & F & 0.401 &\cite{2005ApJ...626...95S} \\ 
2345$-$167  & \n & 23$\mathrm{^h}$48$\mathrm{^m}$02.6085$\mathrm{^s}$ &$-$16$\arcdeg$31$\mathrm{^\prime}$12.0223$\mathrm{^{\prime\prime}}$  & \phantom{0}12 & Q & F & 0.576 &\cite{1993MNRAS.263..999T} \\ 
2351+456  & 4C +45.51 & 23$\mathrm{^h}$54$\mathrm{^m}$21.6802$\mathrm{^s}$ &+45$\arcdeg$53$\mathrm{^\prime}$04.2365$\mathrm{^{\prime\prime}}$  & \phantom{00}8 & Q & F & 1.986 &\cite{1996ApJS..107..541L} \\ 
\enddata

\tablenotetext{a}{Source classified as a quasar in the \cite{VCV06} catalog.}
\tablenotetext{b}{Source classified as a possible BL Lac object in the \cite{VCV06} catalog.}
\tablenotetext{c}{Source classified as a galaxy in the \cite{VCV06} catalog.}
\tablenotetext{d}{Source not listed in the \cite{VCV06} catalog.}
\tablenotetext{e}{Compact steep spectrum source.}

\tablecomments{Columns are as follows: (1) IAU Name (B1950.0); (2) alternate name; (3) right ascension (J2000); (4) declination (J2000);(5) number of 15 GHz VLBA epochs analyzed;   (6) optical classification according to the \cite{VCV06} catalog (with exceptions as noted), where Q = quasar, B = BL Lac object, G = active galaxy, and U = unidentified;  (7) description of radio spectrum from \cite{KL04}, where F=flat, S=steep, and P=peaked; (8) redshift; (9) literature reference for redshift. }

\end{deluxetable*} 

%% file: 2maptablestub.tex
\begin{deluxetable*}{lllccrrrccc} 
\tablecolumns{11} 
\tabletypesize{\scriptsize} 
\tablewidth{0pt}  
\tablecaption{\label{maptable}Summary of 15 GHz Image Parameters}  
\tablehead{ &  & \colhead{VLBA} &\colhead{Freq.} & 
\colhead{$\mathrm{B_{maj}}$} &\colhead{$\mathrm{B_{min}}$} & \colhead{$\mathrm{B_{pa}}$} &  
\colhead{$\mathrm{I_{tot}}$} &  \colhead{rms}  &  \colhead{$\mathrm{I_{base}}$} & \colhead{Fig.} \\ 
\colhead{Source} & \colhead {Epoch} & \colhead{Code} &\colhead{(GHz)} & 
\colhead{(mas)} &\colhead{(mas)} & \colhead{(\arcdeg)} &  
\colhead{(Jy)} & \colhead{(mJy bm$^{-1}$)}  &  \colhead{(mJy bm$^{-1}$)} & \colhead{Num.} \\ 
\colhead{(1)} & \colhead{(2)} & \colhead{(3)} & \colhead{(4)} &  
\colhead{(5)} & \colhead{(6)} & \colhead{(7)} & \colhead{(8)} & \colhead{(9)}& \colhead{(10)} } 
\startdata 
0003$-$066  & 1995 Jul 28 & BZ014\tablenotemark{b} & 15.3 & 1.17 & 0.51 & 0 & 2.172 & 0.5 & 3.0 & 1.1   \\ 
  & 1996 Oct 27 & BK37D\tablenotemark{b} & 15.4 & 1.20 & 0.45 & $-$7 & 1.893 & 0.4 & 0.6 & 1.2   \\ 
  & 1998 Oct 30 & BK52D\tablenotemark{b} & 15.4 & 1.36 & 0.44 & $-$14 & 2.642 & 0.4 & 1.5 & 1.3   \\ 
  & 1999 Jun 17 & BG091 & 15.4 & 1.36 & 0.53 & $-$9 & 2.707 & 0.6 & 1.4 & 1.4   \\ 
  & 2000 Jan 11 & BK68D\tablenotemark{b} & 15.4 & 1.83 & 0.50 & $-$14 & 2.431 & 0.3 & 4.4 & 1.5   \\ 
  & 2001 Jan 21 & BK68H\tablenotemark{b} & 15.4 & 1.62 & 0.58 & $-$12 & 1.878 & 0.3 & 0.9 & 1.6   \\ 
  & 2001 Oct 31 & BR77A\tablenotemark{b} & 15.4 & 1.59 & 0.52 & $-$12 & 1.559 & 0.2 & 1.4 & 1.7   \\ 
  & 2002 May 19 & BG121A & 15.3 & 1.47 & 0.57 & $-$8 & 2.192 & 0.5 & 1.2 & 1.8   \\ 
  & 2003 Feb 5 & BL111E\tablenotemark{a} & 15.4 & 1.32 & 0.53 & $-$6 & 2.847 & 0.2 & 0.7 & 1.9   \\ 
  & 2004 Mar 22 & BG144A & 15.4 & 1.34 & 0.54 & $-$3 & 3.224 & 0.5 & 1.8 & 1.10   \\ 
  & 2004 Jun 11 & BL111M\tablenotemark{a} & 15.4 & 1.30 & 0.49 & $-$7 & 3.302 & 0.3 & 1.5 & 1.11   \\ 
  & 2005 Mar 23 & BL123D\tablenotemark{a} & 15.4 & 1.33 & 0.53 & $-$5 & 3.034 & 0.3 & 0.9 & 1.12   \\ 
  & 2005 Jun 3 & BL123H\tablenotemark{a} & 15.4 & 1.37 & 0.53 & $-$8 & 3.017 & 0.2 & 0.9 & 1.13   \\ 
  & 2005 Sep 16 & BL123M\tablenotemark{a} & 15.4 & 1.33 & 0.52 & $-$6 & 2.718 & 0.2 & 0.5 & 1.14   \\ 
\enddata 
\tablecomments{Columns are as follows: (1) IAU Name (B1950.0); (2) date of VLBA observation; (3) VLBA experiment code; (4) observing frequency in GHz; (5) FWHM major axis of restoring beam (milliarcseconds); (6) FWHM minor axis of restoring beam (milliarcseconds); (7) position angle of major axis of restoring beam; (8) total I flux density (Jy);  (9) rms noise level of image (mJy per beam); (10) lowest I contour (mJy per beam); (11) figure number.}

\tablenotetext{a}{Full polarization MOJAVE VLBA epoch}
\tablenotetext{b}{2 cm VLBA Survey epoch}

\end{deluxetable*} 